\documentclass{aa}  
\usepackage{graphicx}
\usepackage{txfonts}
\begin{document}

\title{Unveiling extremely veiled T Tauri stars
          \thanks{Based
	  on observations collected at the European Southern Observatory and the SMARTS facilities at the Cerro Tololo Inter-American Observatory Observatory in Chile, and the Nordic Optical Telescope, La Palma in Spain.} }


   \author{G. F. Gahm\inst{1}
           \and
	   F. M. Walter\inst{2}
	   \and
	    H. C. Stempels\inst{3}
	   \and
	    P. P. Petrov\inst{4}
	   \and
	   G. J. Herczeg\inst{5}
	   }

   \offprints{G. F. Gahm}

   \institute{Stockholm Observatory, AlbaNova University Centre, Stockholm University,
	      SE-106 91 Stockholm, Sweden\\
              email: \mbox{gahm@astro.su.se}
              \and
	     Department of Physics and Astronomy, Stony Brook University, Stony Brook, NY 11794-3800, USA
	      \and
	      School of Physics \& Astronomy, University of St Andrews, North Haugh, St Andrews KY16 9SS, Scotland
	      \and
	      Crimean Astrophysical Observatory, p/o Nauchny, Crimea, 98409 Ukraine
	      \and 
	      California Institute of Technology, MC 105-24, 1200 East California Boulevard, Pasadena, CA 91125, USA
	      }

   \date{}

   

 \abstract
   {Photospheric absorption lines in classical T Tauri stars (CTTS) are weak compared to normal stars. This so-called veiling is normally identified with an excess continuous emission formed in shock-heated gas at the stellar surface below the accretion streams.} 
   {We have selected four stars (RW Aur A, RU Lup, S CrA NW and S CrA SE) with unusually strong veiling to make a detailed investigation of veiling versus stellar brightness and emission line strengths for comparisons to standard accretion models.}
   {We have monitored the stars photometrically and spectroscopically at several epochs .}
   {In standard accretion models a variable accretion rate will lead to a variable excess emission. Consequently, the stellar brightness should vary accordingly. We find that the veiling of absorption lines in these stars is strongly variable and usually so large that it would require the release of several stellar luminosities of potential energy. At states of very large line dilution, the correspondingly large veiling factors derived correlate only weakly with brightness. Moreover, the emission line strengths violate the expected trend of veiling versus line strength. The veiling can change dramatically in one night, and is not correlated with the phase of the rotation periods found for two stars.}
  {We show that in at least three of the stars, when the veiling becomes high, the photospheric lines become filled-in by line emission, which produces large veiling factors unrelated to changes in any continuous emission from shocked regions. We also consider to what extent extinction by dust and electron scattering in the accretion stream may affect veiling measures in CTTS.  We conclude that the degree of veiling cannot be used as a measure of accretion rates in CTTS with rich emission line spectra.}


 \maketitle
%

\section{Introduction}

Classical T Tauri stars (CTTS) are low-mass pre-main-sequence objects with intense emission line spectra. In standard models the spectacular emission spectrum is explained as the manifestation of magnetospheric accretion of material from a circumstellar disk. Gas is heated while accelerated towards the stellar surface, where energy dissipates in strong shocks. Since the stars may host large-scale organized magnetic fields the shocked gas, producing an excess continuous emission, would be distributed over areas located at high latitudes surrounding the magnetic poles. In addition, the stars show flares and chromospheric/coronal/wind activity. For reviews of properties and models of CTTS see Petrov (\cite{petrov03}) and Bouvier et al. (\cite{bouvier07a}). 

The photospheric absorption lines are diluted (an effect known as veiling). By comparing such spectra with those of non-veiled stars one can derive the energy distribution of the veiling. These distributions have been matched with Paschen continuous emission as well as black body radiation, and for CTTS the derived temperatures are of the order of  10$^{4}$ K (see e.g. Basri \& Batalha \cite{basri90}; Hartigan et al. \cite{hartigan91}; \cite{hartigan95}; Valenti et al. \cite{valenti93}; Calvet \& Gullbring {\cite{calvet98}; Gullbring et al. \cite{gullbring98}). The amount of veiling has been used to estimate accretion rates in TTS. One expects that when the accretion rate changes, the veiling and the stellar brightness also  will change accordingly. However, little systematic, long-term spectroscopic and photometric monitoring of TTS has been performed to explore this prediction. The first studies of this kind, based on low- to medium-resolution spectroscopy, gave the confusing result that although some stars appeared to obey the expected correlation between the degree of veiling and stellar brightness, other stars, and especially some heavily veiled stars, did not (Gahm et al. \cite{gahm95}; Hessman \& Guenther \cite{hessman97}; Chelli et al. \cite{chelli99}). Later high-resolution studies showed that the predicted veiling-brightness relation does not always hold in RW Aur A (Petrov et al. \cite{petrov01}; Petrov \& Kozack \cite{petrov07}), while in AA Tau it does to some extent (Bouvier et al. \cite{bouvier03}, \cite{bouvier07b}). In the present {\it Letter} we investigate the behaviour of veiling in relation to stellar brightness and emission line fluxes in RW Aur A, RU Lup and the binary S CrA (NW and SE). This project is part of a monitoring campaign of CTTS involving optical and NIR spectroscopy, photometry and X-ray  observations.

\section{Observations}

High-resolution optical spectra of RU Lup and S CrA were obtained during consecutive nights in 2000, 2002 and 2005 with the UVES spectrograph on the \mbox{8-m} VLT/UT2 of the European Southern Observatory. For descriptions of the observations and data reductions see Stempels et al. (\cite{st07}, Paper~1). The individual spectra of the S CrA components, separated by only $1\farcs3$, are resolved.  RW Aur A was observed during several nights each year from 1995 to 1999 using the SOFIN spectrograph at the Nordic Optical Telescope, and the observations and reductions were described by Petrov et al. (\cite{petrov01}, Paper 2). 

We have compiled a long-term record of RU Lup and the two S CrA components with $UBVRI/JHK$ photometry since 2003, using the ANDICAM dual-channel imager on the SMARTS 1.3 m telescope at Cerro Tololo. For RW Aur A we derived $V$ magnitudes from stars in the CCD-field at NOT with exposures taken before and after the spectroscopic observations.

\section{Results}

All four program stars are "extreme" CTTS, with very prominent emission line spectra and with only weak traces of photospheric absorption lines, implying that the veiling is high. We have obtained precise measures of the veiling factor (VF), defined by the ratio of the excess continuum to the intrinsic photospheric continuum flux. The VFs were determined by comparing observed spectra with synthetic template spectra and, in the case of RW Aur A, with observed template spectra. The procedures are described by Stempels \& Piskunov (\cite{st03}), where also the veiling of RU Lup is discussed, as well as in Papers 1 and 2. The veiling was measured in spectral windows  containing both visible and isolated absorption lines. This allowed us to derive the energy distributions of the veiling showing, as is normal, a rise towards the blue spectral region. The veiling is strongly variable in all stars.  No star ever fell below VF = 1 in the $V$ band region (corresponding to a 50\% decrease in line strength), and values exceeding VF = 10 were measured. If such high veilings result from excess continuous emission, then the flux in V should become very large, corresponding to several stellar luminosities being liberated at the stellar surface from gravitational energy deposited by the accretion streams. 

\subsection{Veiling versus stellar brightness}

\begin{figure}[t]
\centering
\includegraphics[angle=90, width=6cm]{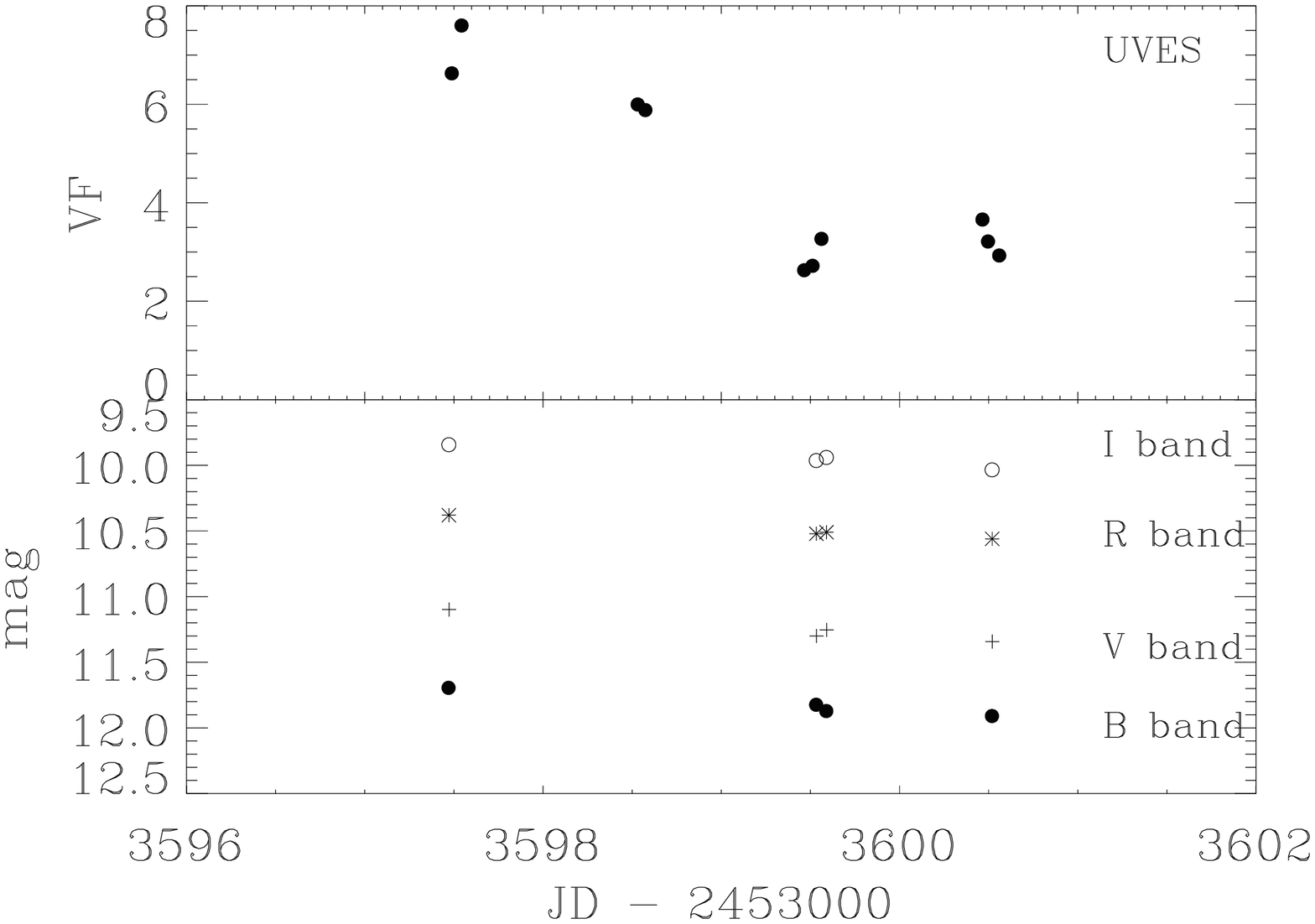}
\includegraphics[angle=00, width=8.5cm]{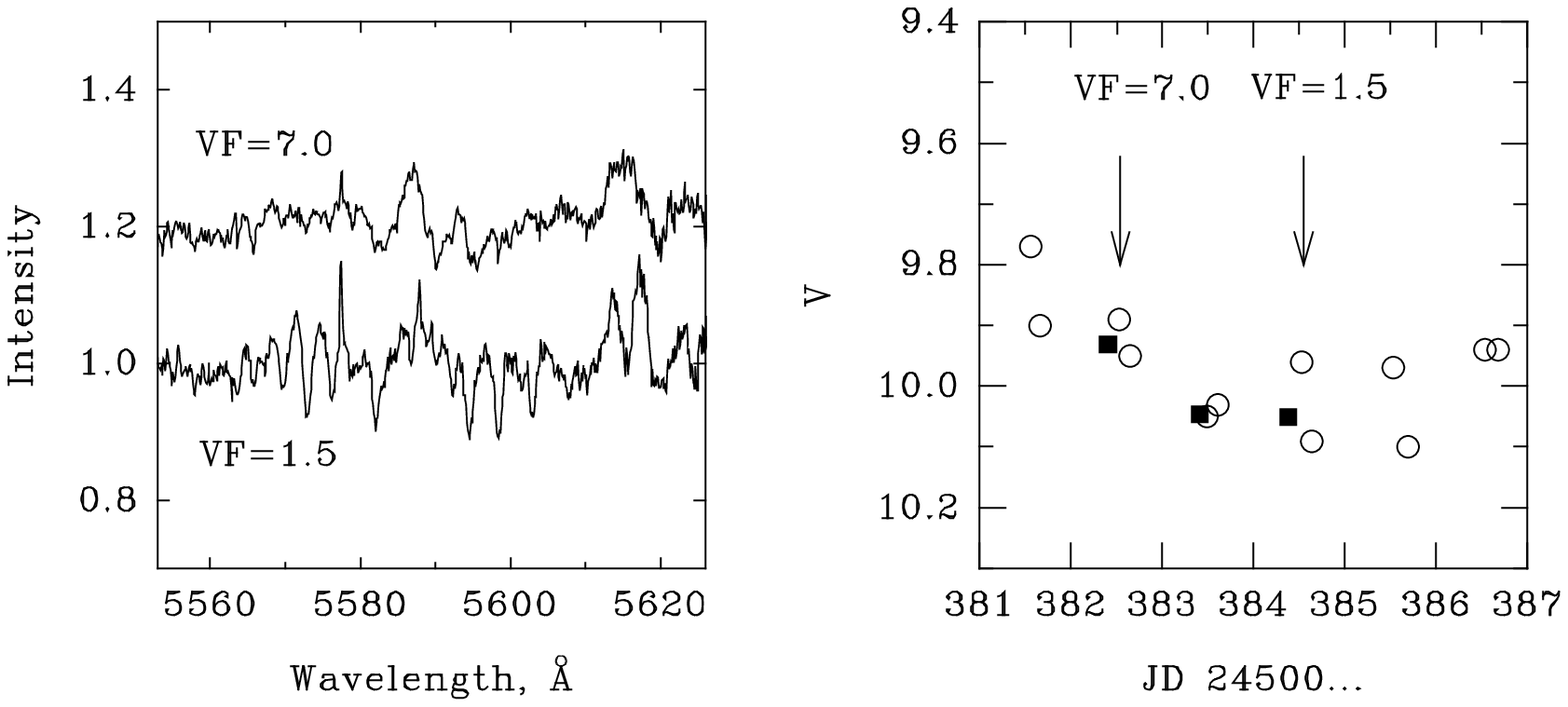}
\caption{{\it Top panels}: Overview of  veiling factors derived from UVES spectra collected over 4 nights of RU Lup in August 2005 including two nights of extremely high veiling, and SMARTS $BVRI$ photometry for the same period. {\it Bottom left}: Spectral region of RW Aur A obtained at high and low degrees of veiling.  {\it Bottom right}: Veiling factor and $V$ magnitude of RW Aur~A during an event where the veiling declined drastically from an extremely high level (circles: our NOT photometry;  squares: Grankin et al. (\cite{grankin07})).}
\label{fig:RURW}
\end{figure}

RU Lup shows the same photometric behaviour as in numerous earlier investigations (see e.g. Giovannelli \cite{giovannelli94} and Pojmanski \cite{pojmanski02}). The star rarely varies by more than $\pm 0.4^{m}$ in $V$ and $\pm 0.15^{m}$ in $I$. SMARTS data collected nearly simultaneously with our spectroscopic observations in 2005 is shown in Fig.~\ref{fig:RURW} (top panels) together with veiling factors for the $V$ band. During the first two spectroscopic nights the veiling was extremely high (VF $> 5$, peaking at 7.6) after which it declined to more 'normal' levels of VF $\approx 2.6$. We recorded a small decline in brightness of $\approx 0.2^{m}$ in all bands over this period. If the changes in veiling were related to a hot area below the accretion stream, then the expected decrease in $V$ should amount to $\approx 1^{m}$. Similarly for RW Aur A (see Fig.~\ref{fig:RURW}, bottom panels), the veiling can decrease from VF = 7.0 to 1.5 over 2 days, while the flux in $V$ decreases by $\approx 0.1^{m}$ compared to 1.3$^{m}$ expected from a fading hot area.
 
Both S CrA components vary dramatically in brightness (Walter \& Miner {\cite{walter05}), and we recorded large changes in the veiling, which on occasion rises to VF $>$ 10.  Photometry and spectroscopy secured in 2005 show that the coupling between veiling and brightness is weak in both components. During this run the VF increased steadily from 5.9 to 10.7 in S CrA NW. This rise was accompanied by a brightening of only 0.1$^{m}$ in $V$, while 0.6$^{m}$ would be expected from a changing hot area. In S CrA SE, when VF rose by 30\% from Aug. 16 to 17, there was no detectable change in $VRI$, but the star became brighter in $B$.  

Hence, in all four program stars, the veiling is unrelated, or only weakly related to brightness during states of {\it high} veiling. {\it These results are inconsistent with the interpretation that veiling as measured from high-resolution spectroscopy is related to the flux level in the accretion continuum.}

For spectra of RU Lup taken in 2002 we could use a secondary flux standard to flux-calibrate our spectra on three nights. This allowed us to calculate synthetic $B$ and $V$ magnitudes of RU Lup, and directly compare the stellar brightness with veiling. Individual measures of magnitude and colour are uncertain but during one night, when VF increased from 2.4 to 4.2, the star brightened in $V$ by $\sim 0.5^{m}$, which is entirely consistent with a brightening of a source of continuous emission on the stellar surface.

\subsection{Veiling versus line flux and phase}

The equivalent widths (EW) of the emission lines in RU Lup vary with time. We have measured EWs for a large number of different types of lines formed in different regions around the star and investigated how EW depends on VF (as measured in the $V$ band). In addition, we have studied how the line intensities correlate with VF across the lines. The method of correlation analysis was first described by Gameiro et al. (\cite{gameiro06}). Fig.~\ref{fig:EWveil} illustrates these dependancies for two lines, namely He {\sc i} ${\lambda}$ 5876 {\AA} and [S {\sc ii}] ${\lambda}$ 6731 {\AA}. The [S {\sc ii}]  line is thought to form in an extended wind, where the red-shifted emission is occulted by the disk, while the central He {\sc i} emission is thought to form close to shocked regions on the stellar surface. Part of the blue wing in He {\sc i} may trace a central stellar wind. Changes in EW are not necessarily related to changes in line flux. In the case of a constant line flux but a variable continuous excess emission the EWs would follow the dashed curves in Fig.~\ref{fig:EWveil}. There are long-term changes in the general level of line emission, but during states of moderate veiling, VF $\la$ 3.5,  the [S {\sc ii}] lines tend to follow the case of constant line flux and varying excess continuum. The He {\sc i} lines show the opposite trend in that during states of low to intermediate veiling EW increases with VF. This is particularly true for the blue-shifted part of the emission, indicating that an increase in the accretion rate/accretion luminosity leads to increased emission from outflowing He {\sc i} gas. However, when the veiling becomes large EW stays constant with increasing VF both for He~{\sc i} and [S~{\sc ii}]. The Balmer lines and emission from metals, like Fe~{\sc i} and Fe {\sc ii}, also saturate in EW for values of VF $\ga$ 3.5. At lower degrees of veiling the metal lines show a larger scatter in EW indicating that intrinsic flux changes occur on shorter time-scales, which will be discussed elsewhere.

We conclude that {\it when the veiling increases to very high levels, the flux in lines formed in very different regions around the star do not respond in the same way}. This provides an independent illustration of our finding that high veiling is unrelated to emission from a shocked region. Both RW Aur A and RU Lup show regular radial velocity changes with periods of a few days (Paper 1 and 2). We did not find any correlation between degree of veiling and phase for these stars. Nor did we find any correlation between $V$ magnitude and phase from the available photometric data banks. 

\begin{figure}[t] 
\centering
\resizebox{7.5cm}{!}{\includegraphics[angle=00]{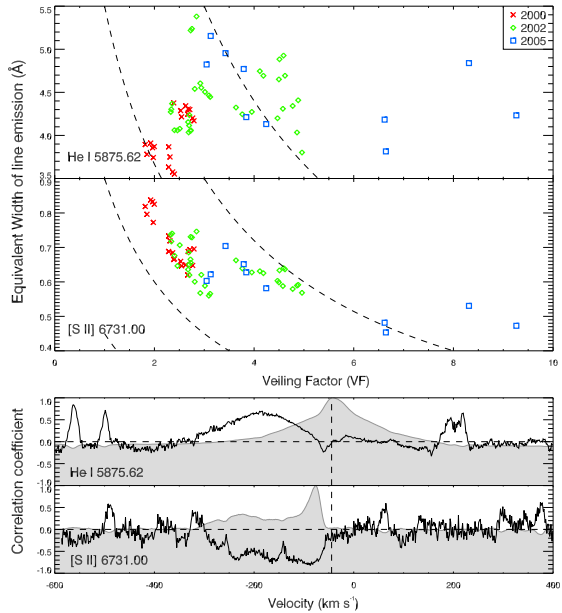}}
\caption{{\it Top panels:} Equivalent width (EW) versus veiling factor (VF), measured in the $V$ band, for emission lines of He {\sc i} and [S {\sc ii}] for RU Lup, where individual runs are marked. The dashed curves correspond to the case of a variable stellar continuum but a constant emission line flux. {\it Bottom panels:} The corresponding correlation functions  together with the average spectrum (shaded), showing how intensity across the line profiles correlates with veiling. The two lines show opposite correlations in the blue wings, but only when VF $\protect\la$ 3.5. The narrow peaks in the correlation functions reflect the behavior of absorption lines not visible in the shaded spectrum. }
\label{fig:EWveil}
\end{figure}

\section{Discussion}

Optical veiling is commonly interpreted as the result of excess continuous emission produced in hot areas surrounding the magnetic poles, where accreting gas is violently halted. The photospheric absorption lines become diluted by this continuum, and consequently the degree of dilution, measured by the veiling factors, should provide a measure of the strength of the emission continuum.  We have found that this prediction may hold true in our sample of rather "extreme" CTTS, with strong metallic line emission, but only for low to moderate values of the veiling factors. During periods when the dilution of absorption lines is very large, the veiling factors are unrelated to any excess continuous emission. This is evidenced from photometric variability versus degree of veiling, and also from the behaviour of emission line strength versus veiling.

When the veiling fluctuates around low to moderate values, which happens frequently in RU Lup, there can be a positive veiling versus brightness correlation in accordance with the standard accretion model, and the equivalent widths of emission lines vary accordingly. Blue-shifted He {\sc i} emission becomes enhanced when the excess emission increases, unlike the forbidden lines arising in an extended wind. The enhanced He {\sc i} emission indicates that a central stellar wind component is excited during an accretion event, while the extended wind component is not affected. However, when the veiling increases to large values the strengths of all different types of emission lines remain constant. 

In AA Tau, the comparatively weak veiling varies periodically with the rotational period (Bouvier et al. \cite{bouvier07b}), and the line strength of  the He {\sc i} emission correlates with the veiling, suggesting it originates at a hot surface region moving with rotation. In RW Aur and RU Lup, we found dramatic changes in the degree of veiling even during one night, but no evidence that the veiling varies with rotational phase. 

Therefore, "extreme" CTTS may host sources of variable continuous excess emission distributed over a small surface as predicted in magnetospheric models. However, additional processes are needed to generate the substantial weakening observed in the photospheric lines. Below we will discuss three possible causes.

{\it A. Variable extinction.} If infalling dust grains survive an accretion event, then the related increase in extinction in the line-of-sight might under specific conditions balance the increase in footprint brightness. This possibility was discussed for AA Tau, with an almost edge-on disk (Bouvier et al. \cite{bouvier03}, \cite{bouvier07b}), and for RW Aur A (Paper 2), which on occasion shows strong inverse P Cygni profiles indicative of events of sudden accretion in the line-of-sight. However, in RW Aur A there is a lack of dependence of strength of accretion components in Na D lines on veiling (Petrov \& Kozack \cite{petrov07}). RU Lup, which is seen nearly pole-on (Herczeg et al. \cite{herczeg05}), has never shown inverse P Cyg profiles. Furthermore, lines formed close to the stellar surface and far out in a wind cannot both be occulted by a local accretion stream. We find that the case of variable extinction in accretion streams is not supported in our study of four CTTS.

{\it B. Electron scattering}. Electron scattering can lead to a dilution of the strengths of photospheric absorption lines without contributing excess continuum emission. This mechanism has not been invoked before in explaining veiling in T Tauri stars, but Stahl \& Wolf ( \cite{stahl80}) discussed e-scattering as a cause for the broad wings of the Balmer lines in S CrA. The characteristic temperature of the infall shock in accreting T Tauri stars is of the order of 10$^5$ K to 10$^6$ K (see e.g., Calvet \& Gullbring \cite{calvet98}). The shocked gas will be highly ionized, so there will be plenty of hot electrons available near the accretion shock. The net effect of scattering of light by hot electrons is to shift the affected photons to shorter wavelength (the inverse Compton effect) and to broaden the frequency distribution (due to the velocity dispersion of the electrons). Code (\cite{code49}) integrated the radiative transfer equation and tabulated the wavelength shift and broadening of a monochromatic line as a function of electron temperature.  The cross section for electron scattering is the Thompson scattering cross section (with minor modifications), so an ionized mass column of about 2.5~g~cm$^{-2}$ above the photosphere will completely erase all underlying spectral information. Once optically thick, electron scattering is a geometric effect, with the veiling dependent on the fraction of the surface covered by this cloud. As an example, if the continuum excess saturates at VF$_0$ = 3, one needs to obscure about a fraction 1-((1+VF$_0$)/(1+VF)) of the star by electron scattering to produce the observed VF. For VF = 4, a screen covering 20\% of the visible surface is needed. Rather special conditions are required for this to occur, such as the hot plasma spreading inside the magnetic funnels after an accretion event has ceased. In particular, it is difficult to envision how the largest values of VF, requiring the largest surface coverage, can build up. Here, we point out that the process as such may be of some importance in TTS, and a more thorough discussion is reserved for a forthcoming paper. As described below, it appears that yet another process affects the veiling measures, and can dominate the dilution of absorption lines in CTTS with strong metallic emission lines.

\begin{figure}[t]
\centering
\includegraphics[angle=00, width=6.5cm]{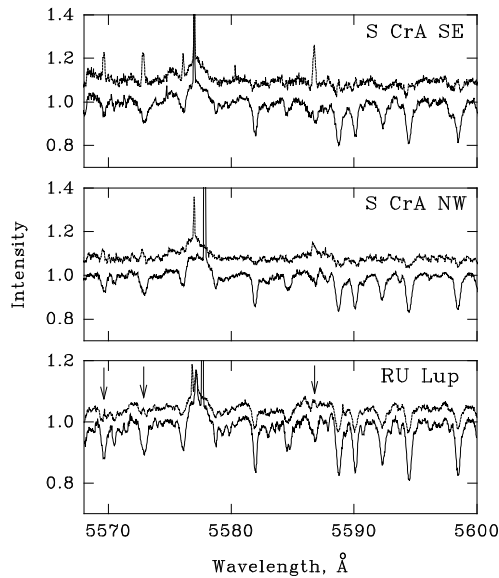}
\caption{Spectral differences for different levels of veiling in S CrA SE and NW, and in RU Lup. A narrow peak of night sky emission enters near 5577 \AA. In the S CrA stars several photospheric lines switch from absorption to emission when the veiling becomes high. Line-filling appears to be present also in certain lines in RU Lup, marked with arrows.}
\label{fig:5580veil}
\end{figure}

 {\it C. Emission reversals in metallic line cores.} In Fig.~\ref{fig:5580veil} we compare the same spectral region obtained at high and low levels of veiling in the S CrA stars and RU Lup. At maximum veiling several lines of Fe {\sc i} are clearly reversed in both S CrA stars, and in RU Lup the same lines appear to be {filled-in by line emission}. No clear dependence of line depth on excitation potential was found for this star, but strong lines are affected most. We conclude that {\it variable narrow emission components filling-in the absorption lines obviously lead to large measured veiling factors without any increase in the continuous excess from any hot surface area}. An increase in line emission will lead to a small increase in brightness. In RW Aur A line-filling is not conspicuous, but may take place (see Fig.~\ref{fig:RURW}).

The emission line components are narrow, indicating a "chromospheric" origin, and that the atmospheric structure changes with time. Similar phenomena are seen in other types of objects, and in the case of CTTS it was once demonstrated by Cram (\cite{cram79}) and Calvet et al. (\cite{calvet84}) how unusual spectral emission signatures could arise from changes in the T -- $\tau$ relation in the stellar atmosphere when the temperature minimum falls below $\tau$ = 1 (denoted the "deep chromosphere" by Herbig (\cite{herbig70})). We found in Sec. 3 that even at high veilings there is often a weak correlation between brightness and veiling. These variations may relate to minor changes in the accretion rate, which in turn could affect the stellar atmosphere surrounding the accretion shock leading to the large veiling factors discussed. We have found periodic velocity changes in some narrow emission components, and the case of variable amounts of chromospheric emission surrounding the magnetic poles will be further explored in a separate paper.

\section{Conclusions}

Our spectroscopic and photometric observations of four classical T Tauri stars with very strong emission signatures show that the degree of veiling is variable, but when the veiling becomes strong it is only weakly dependent on the stellar brightness. Measured emission line strengths support this conclusion. We demonstrate that veiling can be caused by narrow emission lines that fill in the stellar absorption lines and not by a continuous source of emission, as is usually postulated. We discuss two other effects
leading to a reduction of absorption line strengths, and which may play some role in CTTS.  We conclude that the degree of veiling, as measured from absorption lines, leads to an overestimation of the accretion rates in CTTS with rich emission line spectra, especially when the veiling factors VF exceed {$\sim 3$}, as happens frequently in the stars investigated. In extreme cases the overestimation may amount to an order of magnitude.  

\begin{acknowledgements}

This work was supported by the INTAS grant 03-51-6311 and the Swedish National Space Board.

\end{acknowledgements}

\end{document}